\pdfoutput=1
\documentclass[twocolumn,nofootinbib]{revtex4-1}
\usepackage[utf8]{inputenc}
\usepackage[T1]{fontenc}
\usepackage[english]{babel}
\usepackage{graphicx,amssymb,amsmath,hyperref}

\begin{document}

\title{A fast map-making preconditioner for regular scanning patterns}
\author{Sigurd K. Næss}
\email{sigurd.naess@astro.ox.ac.uk}
\author{Thibaut Louis}
\email{thibaut.louis@astro.ox.ac.uk}
\affiliation{Sub-department of Astrophysics, University of Oxford, Keble Road, Oxford, OX1 3RH, UK}

\begin{abstract}
	High-resolution Maximum Likelihood map-making of the Cosmic
	Microwave Background is usually performed
	using Conjugate Gradients with a preconditioner that ignores noise
	correlations. We here present a new preconditioner that approximates
	the map noise covariance as circulant, and show that this results in
	a speedup of up to 400\% for a realistic scanning pattern from the
	Atacama Cosmology Telescope. The
	improvement is especially large for polarized maps.
\end{abstract}

\maketitle

\section{Introduction}
As the resolution and sensitivity of Cosmic Microwave Background (CMB) experiments
increase, so do the computational resources needed to analyze their data.
Because modern detectors are background-limited, the only way to significantly
increase sensitivity is to increase the number of detectors. The last
decades have seen an increase from tens of detectors to thousands of detectors in
experiments like ACT \cite{act-maps,actpol-desc}, SPT \cite{spt-desc,sptpol-desc},
POLARBEAR \cite{polarbear-desc} and Keck \cite{keck-array}, and plans already exist
for experiments with $10^5-10^6$ detectors \cite{cmb-neutrino-s4}. Reducing the data
from all these detectors into a coherent map of the sky presents a
significant computational challenge,
and already with 1000-detector-class experiments this
step is the most important bottleneck of the data analysis pipeline \cite{act-maps}.
It is therefore important to investigate ways to speed up this process.

Three main classes of map-makers are in popular use: \emph{Maximum likelihood map-makers}
\cite{mapmaking-tegmark,madmap2010,roma,romagal,act-maps,quiet-wband},
which are slow, but produce unbiased, optimally noise-weighted maps;
faster but slightly less accurate \emph{destripers} \cite{madam,descart};
and biased and sub-optimal but very fast \emph{naive map-makers} \cite{quiet-wband,spt-maps-2011}.
The topic of this paper is a method for significantly speeding up
maximum likelihood map-makers.

Assuming a linear detector response, we can model the time-ordered
data $d$ via the the linear system
\begin{align}
	d = Pm + n,
\end{align}
where $m$ is the pixelized map of the sky, the pointing matrix $P$
is a sparse\footnote{The pointing matrix will be sparse if we solve
	for a {\em beam-convolved map}. For variable beams or asymmetric
	beams, one may want to reconvolve to a standard beam as part of
	map-making. This can be done using $P$, at the cost of some of its
sparsity. We do not consider this case here.} matrix mapping from pixels to samples, and $n$ is
the time-domain noise, which we assume to be gaussian with covariance
matrix $N$.
The maximum likelihood solution for $m$ is
given by the map-making equation \cite{mapmaking-tegmark},
\begin{align}
	(P^T N^{-1} P) m &= P^T N^{-1} d.
\end{align}
This is of the form $Ax=b$, and while the matrices involved are
usually
too large to solve by direct inversion, the system is amenable to solution by
Preconditioned Conjugate Gradients (PCG) \cite{numerical-recipes}
provided a good preconditioner
can be found\footnote{Without a preconditioner, the number of iterations
needed for conjugate gradients is proportional to the condition number of
the matrix $A$. By applying a preconditioner $M$, one is effectively
solving the system $MAx=Mb$. The goal is then to choose $M$ such that
$MA$ is as well-conditioned as possible. This can be acchevied if
$M \approx A^{-1}$.}. The most commonly used map-making
preconditioners are the binned \cite{first-pcg-map,preplanck-ashdown,madmap2010}
and Jacobi \cite{mapcumba2001,roma,madmap2010,act-maps} preconditioners.
The binned preconditioner $M_\textrm{B}$
approximates the time covariance matrix $N$ as diagonal
(i.e. it ignores correlations), which for pointing
matrices where only one pixel is hit per sample results
in a diagonal pixel-space covariance matrix.
\begin{align}
	N^{-1}_{ij} &\approx N^{-1}_{ii} \delta_{ij} \textrm{ and } P_{ti}P_{tj} \propto \delta_{ij} \Rightarrow \\
	A_{ij} &\equiv P_{ti} N^{-1}_{tt'} P_{t'j} \approx P_{ti}^2 N^{-1}_{tt} \delta_{ij} \equiv {M_\textrm{B}}^{-1}_{ij} \label{eq:binned}
\end{align}
The Jacobi preconditioner $M_\textrm{J}$ simplifies one step further, and assumes that every
detector has the same variance, $N = aI$, resulting in
\begin{align}
	A_{ij} &\approx a P_{ti}^2 \delta_{ij} \equiv {M_\textrm{J}}^{-1}_{ij}.
\end{align}
The proportionality factor $a$ is usually set to 1, as PCG is
insensitive to an overall scaling of the preconditioner.

\begin{figure}
	\centering
	\includegraphics[width=0.4\columnwidth,angle=90]{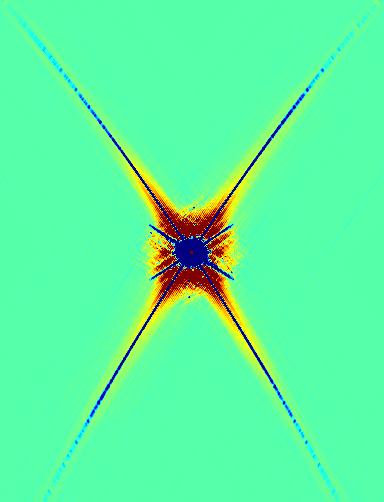}\\
	$-10^{-4}$ \includegraphics[width=0.3\columnwidth,height=0.6em]{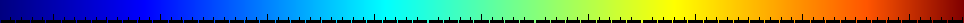} $10^{-4}$
	\caption{A $4.2^\circ$ by $3.2^\circ$ subset of a row from the pixel-space inverse correlation
		matrix from a patch from ACT (each row in the matrix corresponds to a two-dimensional map).
		In order to highlight the
		correlation structure, the color scale is capped at $\pm 10^{-4}$.
		The significantly correlated area has a quite complicated shape,
		which is driven by the scanning pattern and focal plane layout.}
	\label{corr_ex}
\end{figure}

\begin{figure}
	\centering
	\includegraphics[width=0.9\columnwidth]{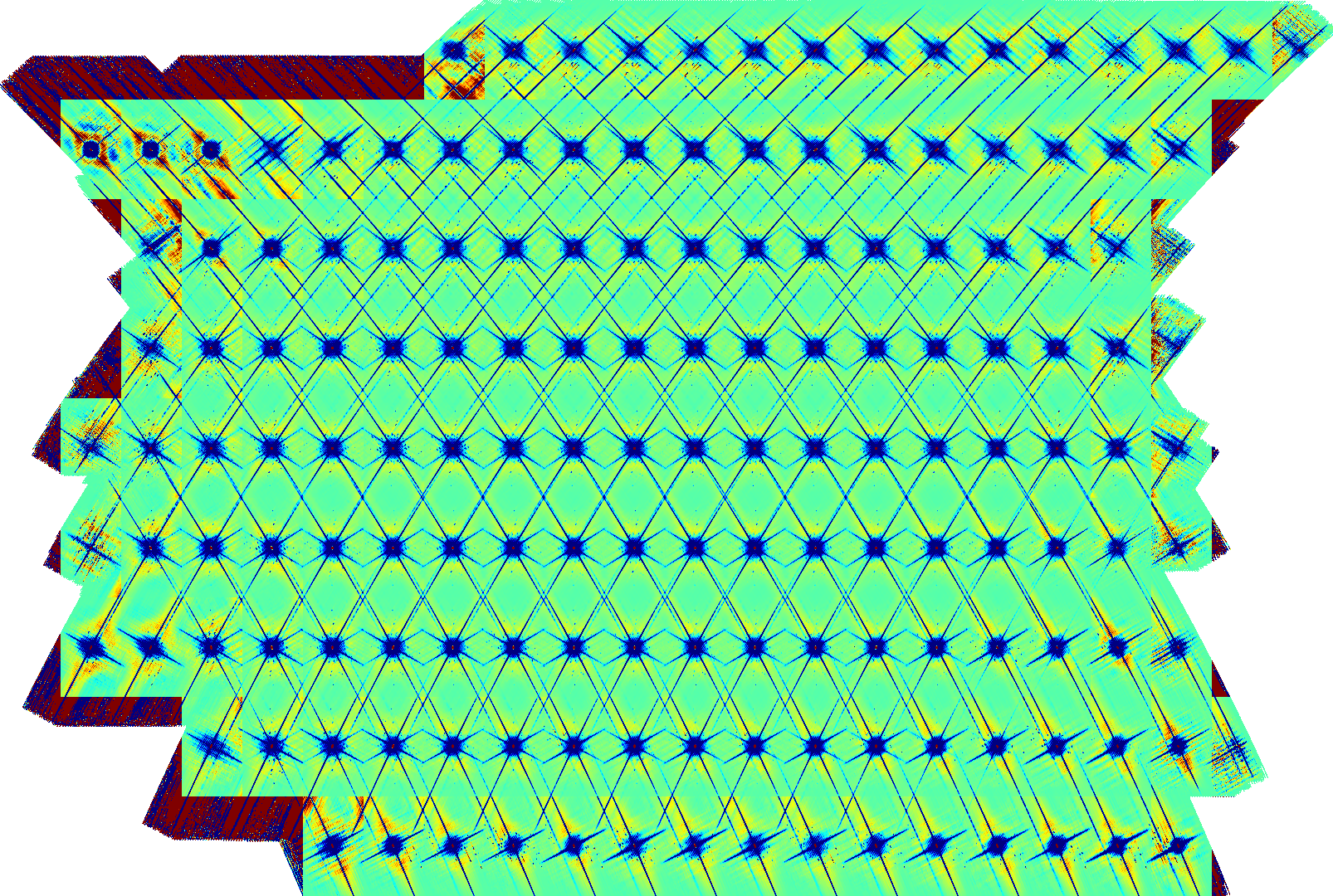} \\
	$-10^{-3}$ \includegraphics[width=0.7\columnwidth,height=0.6em]{figs/colorbar.png} $10^{-3}$
	\caption{A mosaic of the local correlation structure for a $12^\circ$ by $9^\circ$
		subset of an ACT patch. A set of evenly spaced pixels were chosen as reference
		points, and for each the correlation in a neighborhood around it was
		calculated. Each pixel in the map is colored according to its correlation
		relative to the nearest reference point. The correlation structure
		is very uniform in the whole central region of the map.}
	\label{corr_multi}
\end{figure}

The assumption of independent noise is quite inaccurate. All
realistic experiments have at least some time-correlation in
the noise, and usually also correlations between different detectors.
Additionally, filters will also generally introduce correlations.

While computing the full, exact $A$ is often too expensive, single
rows of it can be computed at the same expense as one CG iteration:
\begin{align}
	A_{ij} &= A_{ik} \delta_{kj} = (A \vec e_i)_j
\end{align}
Here $(\vec e_i)_j = \delta_{ij}$ is the pixel-space basis vector corresponding
to pixel $i$. An example of what such a row looks like for the Atacama Cosmology Telescope
(ACT) \cite{act-maps} can be seen in figure~\ref{corr_ex}.

It is clear that the approximation of no correlation is
quite inaccurate. However, since the correlation structure is
driven by the scanning pattern and relative position of the detectors
in the focalplane, the correlation structure should be the same
for all pixels which are scanned the same way. ACT, which
used long-duration, small-amplitude drift scans,
pixels at the same declination but different right ascension
will be hit by the same phase of the same scanning motion, and
should therefore have the same correlation structure.\footnote{
	Circulant correlation is not a good approximation for every experiment.
	It is suitable for constant elevation dift scans, as employed by
	ACT, SPT and POLARBear, but we expect it to work poorly for full-sky
scanning patterns.}

Indeed, that is what measurements show (see figure~\ref{corr_multi}).
It may therefore be a good approximation to assume that every point in
the map has the same relative correlation structure, i.e. that the
correlation between two points on the sky only depends on their relative position.
If this is the case, then it is possible to choose a pixelization where the
correlation only depends on the difference between pixel numbers,
and hence that the pixel correlation matrix is \emph{circulant}\footnote{
	For example, if $\textrm{corr}(\vec x_1,\vec x_2) = f(\vec x_1-\vec x_2)$,
	where $\vec x$ are coordinates,
	then a pixelization scheme $\vec x = G(p)$, where $p$
	is a pixel index and G is a linear function will fulfill
	$\textrm{corr}(p_1,p_2) = f(G(p_1-p_2))$, resulting
in a circulant correlation matrix.}.

A circulant matrix has the nice property of being diagonal in
the frequency domain, which means that it can be computed, stored
and applied cheaply, at the cost of a few
FFTs. Hence, the constant correlation approximation is promising
as a preconditioner for solving the map-making equation.

\section{Implementation}
The inverse pixel covariance matrix $A$ can be decomposed into
variance and correlation such that $A = \Sigma^T U\Sigma$.
Here $\Sigma$ is diagonal (block-diagonal in the case of polarization)
in pixel space, and corresponds to a
map of the inverse standard deviation per pixel. As per the
binned preconditioner, this can be approximated as
\begin{align}
	(\Sigma^T\Sigma)_{ij} &= P_{ti}^2 N^{-1}_{tt} \delta_{ij} .
\end{align}
The correlation matrix $U$ is in general a dense matrix, but
as noted above, it can often be approximated as circulant\footnote{%
	The matrix will be circulant provided that the correlation
	structure is position-independent, that a constant offset in
	each coordinate corresponds to a constant pixel offset, and
	provided that indices wrap around at the edges.}.
Therefore, the constant correlation preconditioner replaces
$U$ with a circulant matrix $Q$, such that
\begin{align}
A &\approx \Sigma^T Q \Sigma \equiv {M_\textrm{C}}^{-1}
\end{align}
This relation can be inverted to give us an expression for $Q$
in terms of $A$,
\begin{align}
	Q =& {\Sigma^T}^{-1} A \Sigma^{-1} \Rightarrow \\
	Q_{ij} =& {\Sigma^T}^{-1}_{ii} (A \vec e_i)_j \Sigma^{-1}_{jj}. \label{eq:Q}
\end{align}
Since $Q$ is circulant, i.e. $Q_{ij} = Q_{0,j-i} = q_{j-i}$, we have
\begin{align}
	(FQF^{-1})_{ff'} &= (Fq)_{-f}\delta_{ff'} = (Fq)^*_f\delta_{ff'}
\end{align}
using forward and backwards Fourier transforms $F_{fj} \equiv
\mathrm{e}^{-\frac{2\pi ijf}{N}}$ and $F^{-1}_{jf} \equiv
\frac1N \mathrm{e}^{\frac{2\pi ijf}{N}}$, where $N$ is the
number of rows in the matrix.

With this in hand, the preconditioner can be applied as
\begin{align}
	M_\textrm{C}b
		&= \Sigma^{-1} Q^{-1} {\Sigma^{-1}}^T b \notag \\
		&= \Sigma^{-1}F^{-1}{(Fq)^*}^{-1}F{\Sigma^{-1}}^T b.
\end{align}
$\Sigma^{-1}$ and ${(Fq)^*}^{-1}$ can be precomputed,
so the cost of applying the preconditioner is simply that of
two FFTs and three diagonal matrix multiplications.

The choice of the reference pixel at which the correlation is
measured is somewhat arbitrary. We used the pixel nearest the
center of the map, but other locations not too close to the
edge of the map should also work.

\begin{figure}[htb]
	\centering
	\includegraphics[width=\columnwidth]{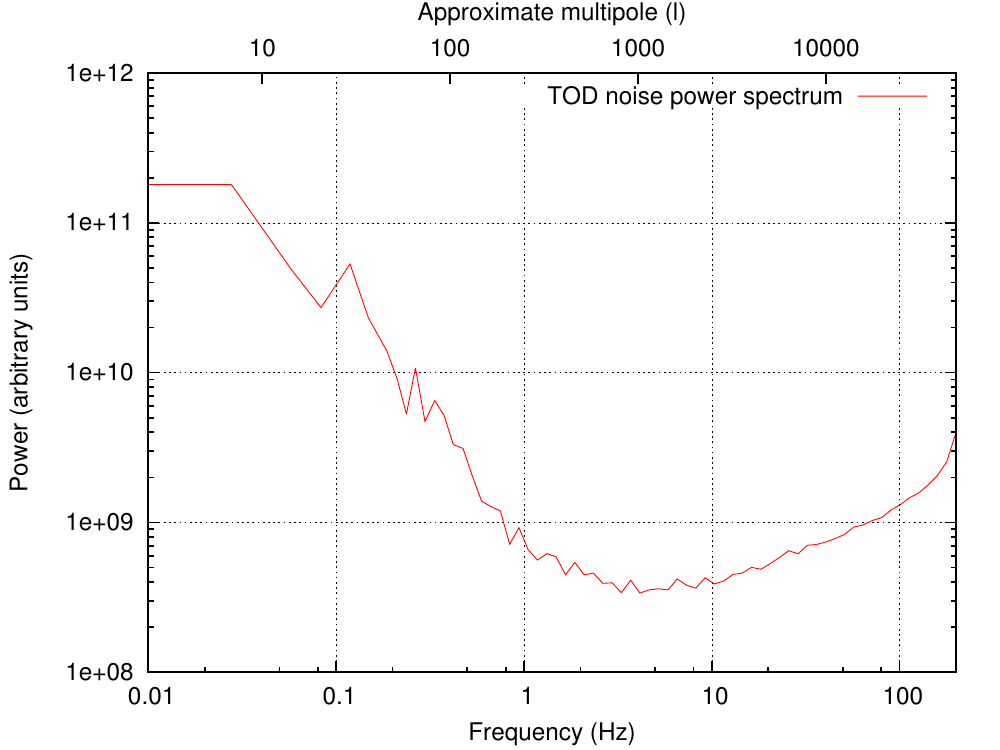}
	\caption{Example TOD noise power spectrum from the simulation, based on
noise behavior from ACT. Low
frequencies are dominated by atmospheric noise, while the increase at
high frequencies is caused by a Butterworth filter.}
	\label{fig:noise}
\end{figure}

\section{Regularization}
While constant correlation is a good approximation for
relatively short-scale correlations, it works less well for
long-distance correlations, and regions near the edges\footnote{%
Near the edges the telescope must decelerate in order to reverse
the scanning direction, which makes the correlation structure
different there than in the center.}
Applying the preconditioner as described above to realistic cases
results in the appearance of large scale modes which change
extremely slowly during the subsequent CG iteration.

A way around this is to artificially limit the range of the correlations
that are modeled, by multiplying $q$ by a Gaussian. For ACT,
a standard deviation of 20 pixels was found to be effective, but
this will depend on the scanning pattern, and
some experimentation may be needed to find the optimal number.

\begin{figure}[htb]
	\centering
	\includegraphics[width=\columnwidth]{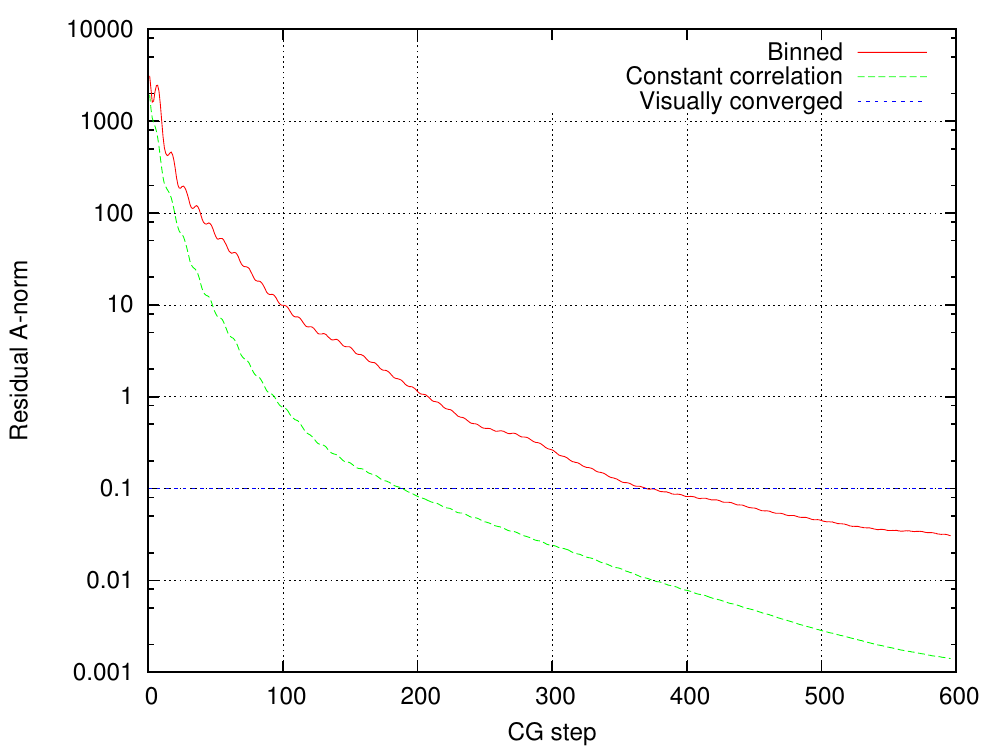}
	\caption{The residual A-norm \cite{cg-errors} as a function of iteration number for
		a simple binned preconditioner and the constant correlation
		preconditioner. The latter converges roughly twice as fast as
		the former according to this criterion. The horizontal blue line
		indicates the level at which the maps have mostly stopped
		changing visually.}
	\label{fig:a-norm}
\end{figure}

\section{Polarization}
The previous discussion assumed that each pixel only had a single
degree of freedom, e.g. temperature-only maps of the sky. In the
case of polarization, each pixel has several correlated components,
typically the Stokes parameters T, Q and U \cite{polarization-basics}, but this can instead
be expressed as a larger
number of block-correlated single-component pixels. This results
in $\Sigma$ being block-diagonal with e.g. one (T,Q,U)-block per physical
pixel, while $Fq$
becomes a vector of similar blocks. And instead of a single row of $A$
needing to be measured, all the rows corresponding to a given physical
pixel now need to be computed (i.e. $\vec e_{i\alpha}$ for all components,
where Greek indices indicate polarization components).
Hence, eq.~(\ref{eq:Q}) becomes
\begin{align}
	Q_{i\alpha j\beta} =& {\Sigma^T}^{-1}_{i\alpha i\gamma} (A \vec e_{i\gamma})_{j\delta} \Sigma^{-1}_{j\delta j\beta}
\end{align}
Aside from that, everything works the same.

\section{Test setup}

\begin{figure*}[htb]
	\centering
	\begin{tabular}{cccccc}
		& Step 5 & Step 15 & Step 45 & Step 115 & Step 340 \\
		\rotatebox{90}{\hspace{11mm}Binned}&
		\includegraphics[width=0.18\textwidth]{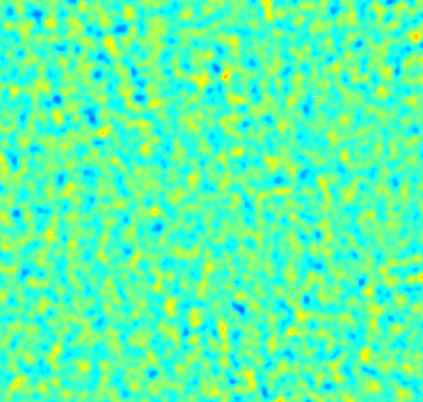} &
		\includegraphics[width=0.18\textwidth]{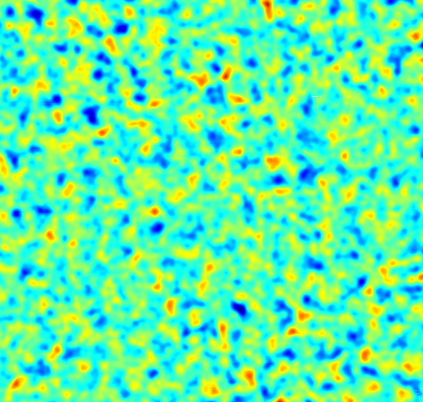} &
		\includegraphics[width=0.18\textwidth]{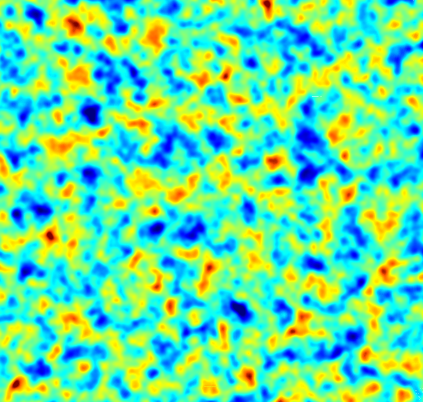} &
		\includegraphics[width=0.18\textwidth]{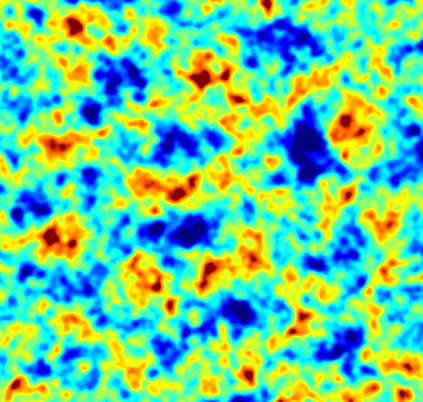} &
		\includegraphics[width=0.18\textwidth]{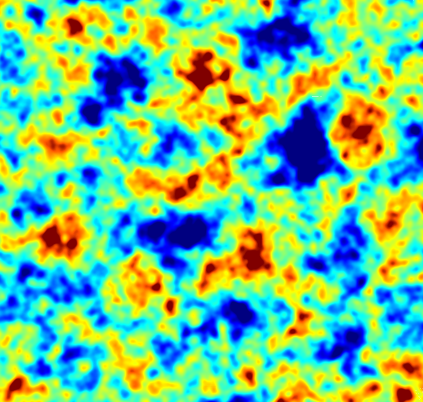} \\

		\rotatebox{90}{\hspace{8mm}Const. corr.}&
		\includegraphics[width=0.18\textwidth]{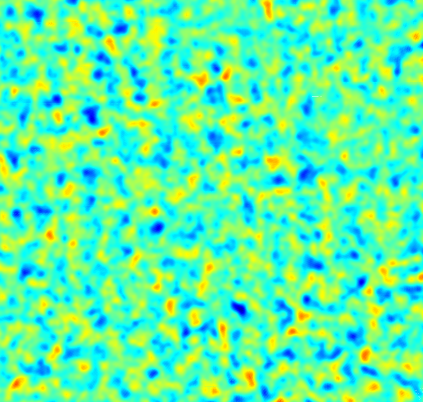} &
		\includegraphics[width=0.18\textwidth]{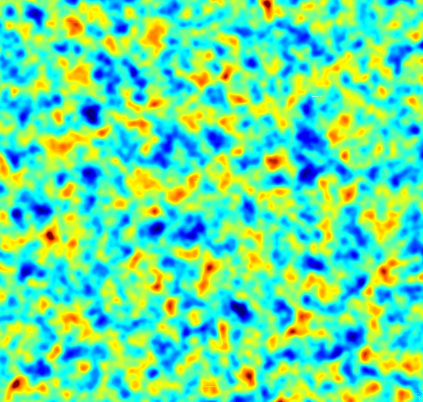} &
		\includegraphics[width=0.18\textwidth]{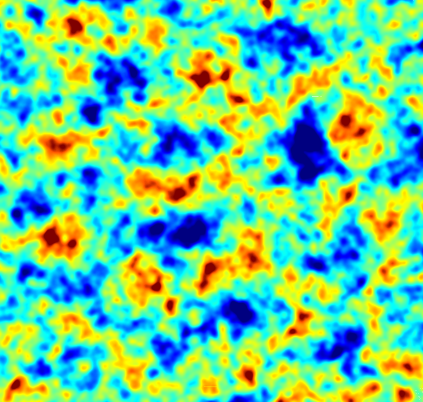} &
		\includegraphics[width=0.18\textwidth]{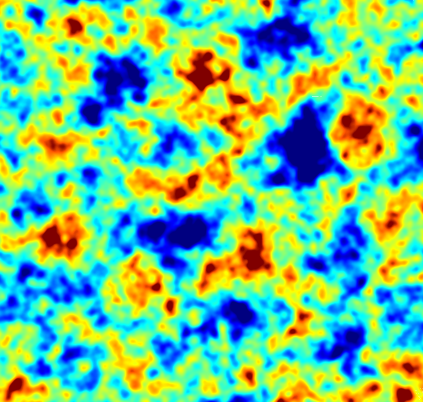} &
		\includegraphics[width=0.18\textwidth]{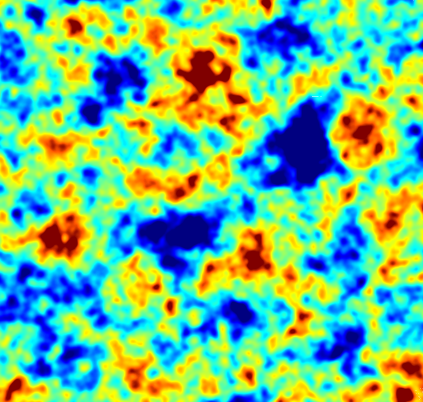}
	\end{tabular}
	\caption{Example temperature maps from the CG solution process for the
		binned (top) and constant correlation (bottom) preconditioners. The
		rows correspond to steps 5, 15, 45, 115 and 340 from left to right.
		The steps are chosen such that the binned map in column $n$ is as similar
		as possible as the constant correlation map in column $n-1$. We see that
		the constant correlation preconditioner visually converges about 3
		times faster than the binned one. The maps have been cropped for compactness
		of presentation.}
	\label{fig:conv_map_t}
\end{figure*}

\begin{figure*}[htb]
	\centering
	\hspace{-10mm}
	\begin{tabular}{ccc}
		Step 15 & Step 70 & Step 250 \\
		\includegraphics[height=0.32\textwidth,clip=true,trim=0mm 0mm 6mm 0mm]{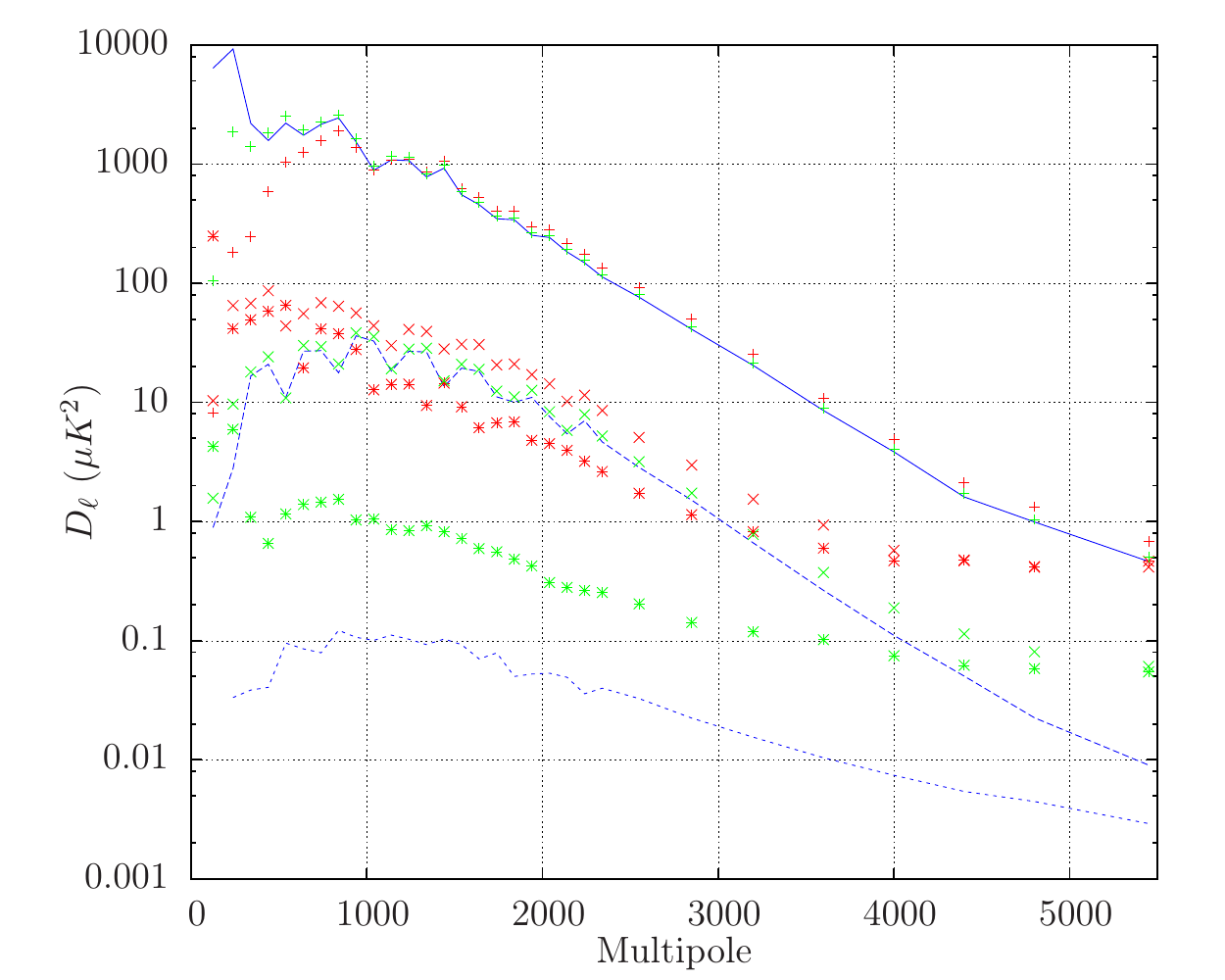} &
		\includegraphics[height=0.32\textwidth,clip=true,trim=19mm 0mm 6mm 0mm]{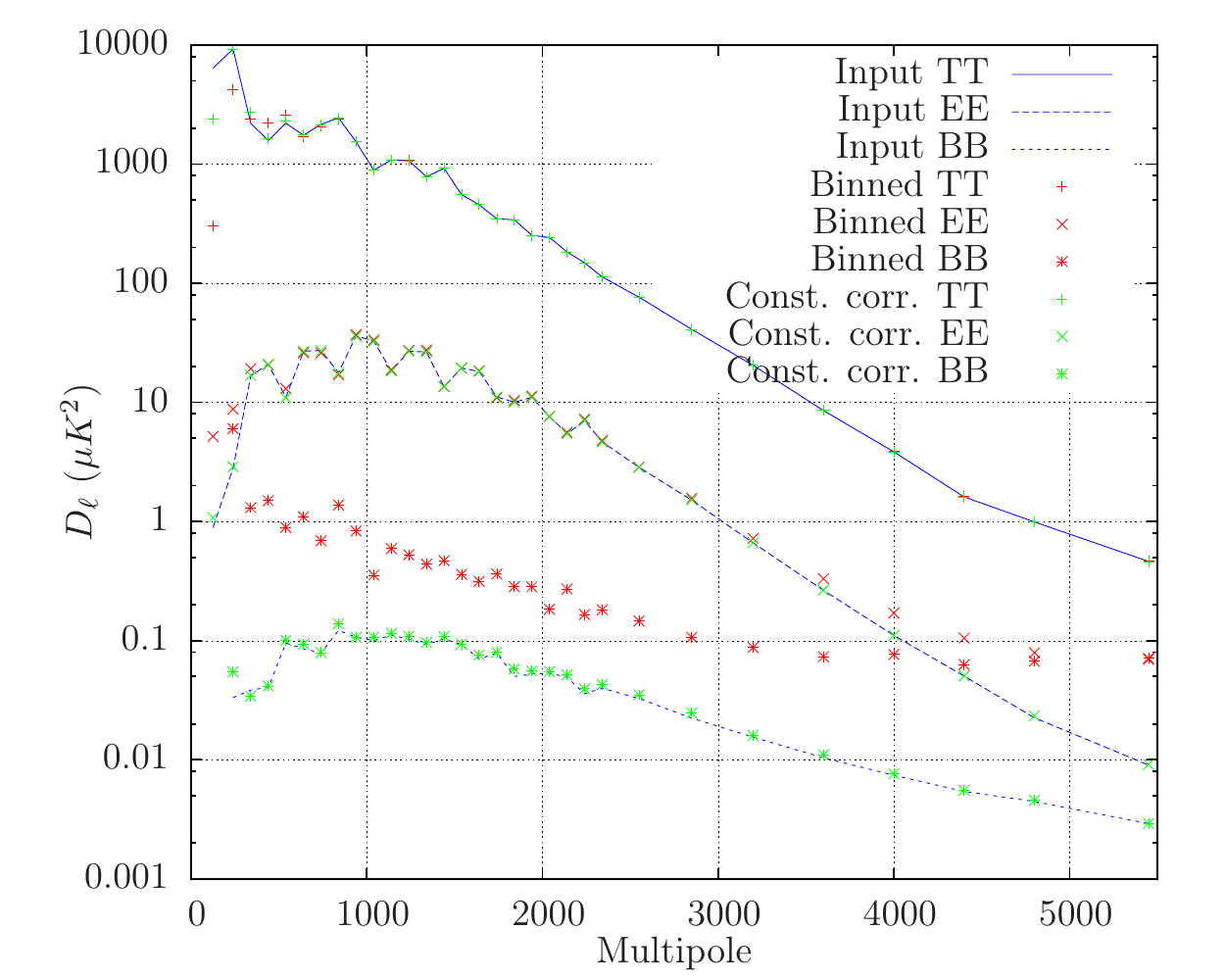} &
	\includegraphics[height=0.32\textwidth,clip=true,trim=19mm 0mm 6mm 0mm]{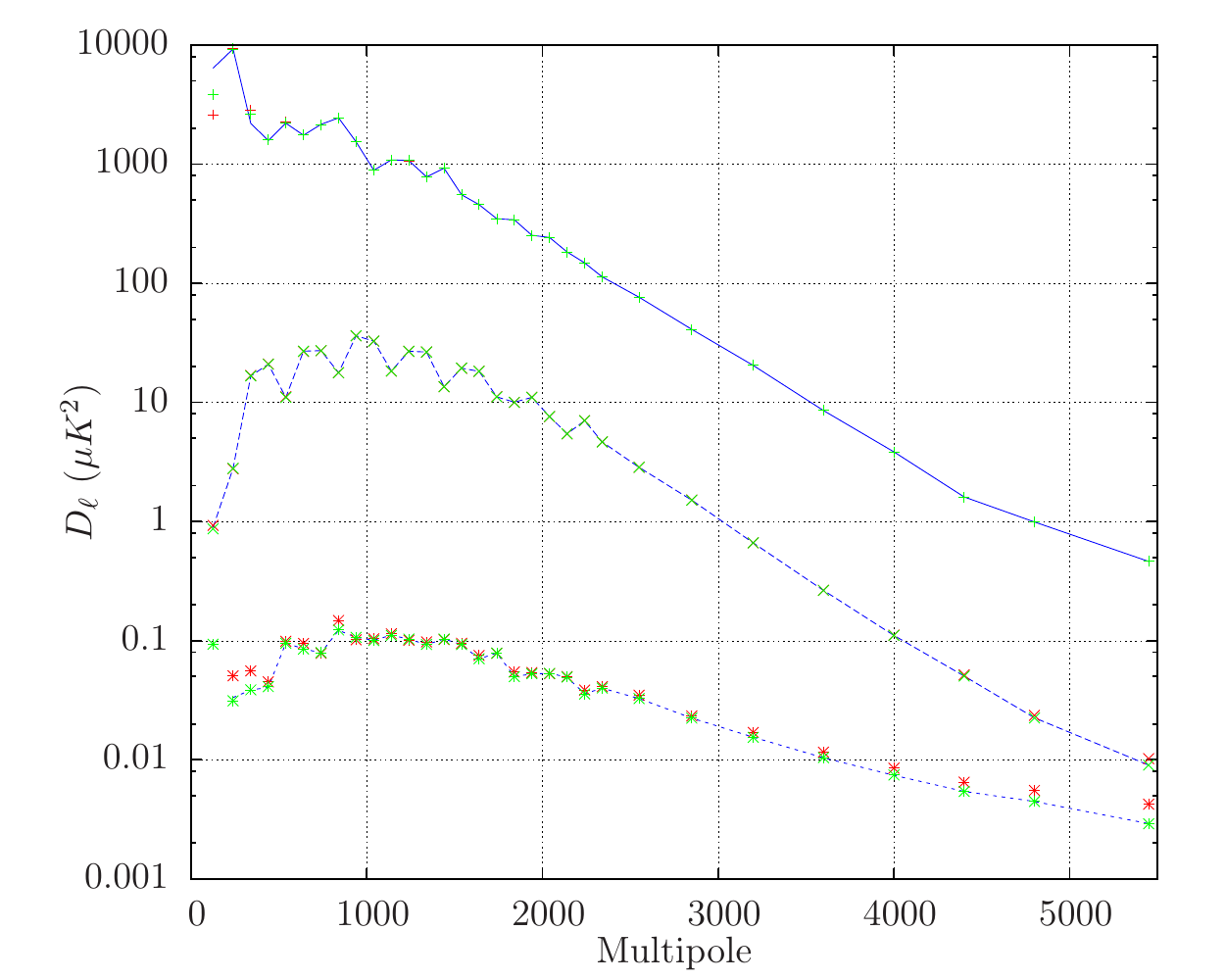}
	\end{tabular}
	\caption{Comparison of the input (blue), binned (red) and constant
		correlation (green) power spectra at CG steps 15 (left), 70 (middle)
		and 250 (right). Unlike the map-space plots in figure~\ref{fig:conv_map_t},
		where only the large-scale convergence is visible, we can here clearly
		see the convergence at all scales. Small-scale convergence is much
		slower for EE and BB than for TT, and form the bottleneck for the
		CG solver if we ignore the $\ell<500$ modes. The constant correlation
		preconditioner is 3-5 times faster than the binned one here. This is
		quantified more precisely in figure~\ref{fig:speedup}.}
	\label{fig:absplot}
\end{figure*}
We tested the preconditioners on a simulated time-ordered data (TOD) based
on the scanning pattern for 64 detectors from each of 417 15-minute
scans of a subset of ACT's southern patch centered at $\alpha=56^\circ,\delta=-53^\circ$.
Each scan was a constant elevation drift scan with amplitude of $3.5^\circ$
in azimuth, and the scan centers were spread over 10 steps in elevation,
covering a patch of about $11^\circ$ by $8^\circ$ degrees.
Odd steps in elevation scanned while rising and the even ones when
setting. This resulted in most pixels being hit from two directions,
and hence the x-shaped correlation pattern seen in figures~\ref{corr_ex}-\ref{corr_multi}.
An example of a noise power spectrum used in the simulation can be seen
in figure~\ref{fig:noise}.

The simulated
detectors were polarization-sensitive, with each detector measuring
a linear combination $T + \cos(2\psi)Q + \sin(2\psi)U$ of the local
radiation field, with each detector having a different, randomly chosen
detector angle $\psi$. While an ACT-like noise model, including the effects
of atmosphere and inter-detector noise correlations was assumed in the
map-making step, no noise was added to the simulated TOD in order to
allow the convergence to be studied all the way to the highest multipoles\footnote{
This is valid since the convergence rate of PCG is mostly independent of the
noise level of the right-hand side after the first few iterations\cite{first-pcg-map}.
However, with higher noise, higher CG errors also become acceptable, so the
number of iterations needed for CG errors to be subdominant will be smaller
for realistic noise levels.}.
For the same reason, the simulated input CMB did not include a beam,
and was pixelated at the same resolution as the output map, in order to
avoid subpixel noise.

We then solved the map-making equation for this data set using PCG, first using
the binned preconditioner described in equation~(\ref{eq:binned}), and then
the constant correlation approximation described in this paper.
Each was run for 600 CG iterations, with intermediate maps being
output for every 5 steps.

\section{Results}

The constant correlation preconditioner visually converges roughly
3 times faster than the binned one, as shown for the temperature
map in figure~\ref{fig:conv_map_t}. Likewise, the residual A-norm \cite{cg-errors} from
the CG solver (shown in figure~\ref{fig:a-norm}) also shows a significant improvement in convergence:
roughly a factor 2 according to this metric\footnote{The A-norm $||x||_A$ of a
	vector $x$ is defined as $\sqrt{x^TAx}$, where $A$ is $P^TN^{-1}P$
	in our case. The error A-norm after $i$ CG steps is $||x_i-x||_A$, where $x_i$
	is our estimate after $i$ steps, and $x$ is the true map. When
	the true $x$ is unknown, $||x_i-x||_A$ can be estimated as $||x_i-x||_A
	\approx \sum_{j=i}^{i+d-1} \gamma_j||r_j||^2$, where $r_j$ and $\gamma_j$
	are two internal variables in the CG algorithm at step $j$, and $d$
	is is an integer that controls the accuracy of the estimate (4 in our case).
}

\begin{figure}[htb]
	\centering
	\includegraphics[width=\columnwidth]{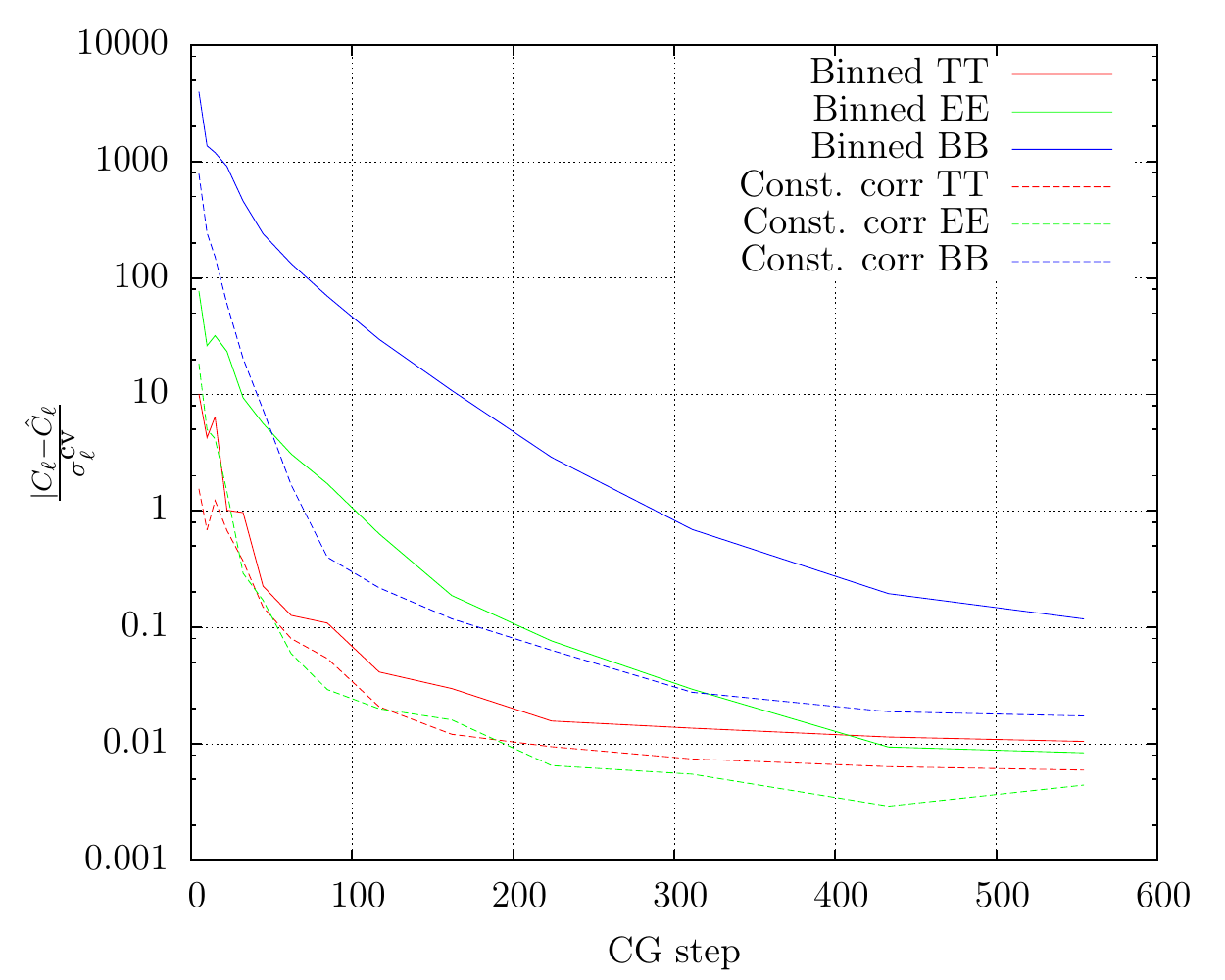}
	\caption{An example of the convergence of a single bin in the power spectrum,
		in this case that centered on $\ell=3200$. This shows the absolute difference
		between the recovered spectrum $C_\ell$ and the spectrum $\hat C_\ell$ of the
		input map in units of cosmic variance $\sigma_\ell^\textrm{cv}$,
		as a function of the CG step in cosmic variance units. Shown are
		curves for TT (red), EE (green) and BB (blue) for the binned (solid)
		and constant correlation (dashed) preconditioners. The trend lines
		are based on binned averages of the errors for many individual CG steps.
		This binning reduces the jitter,
		and makes the trend lines suitable for measuring the time needed to converge
		to a given level.}
	\label{fig:ps_conv}
\end{figure}

However, neither of these tests take into account the fact that not all scales
in the map are equally interesting. To remedy this, figure~\ref{fig:absplot}
compares the binned and constant correlation power spectra with that of the
simulated input map. These spectra were computed using the method
described in \cite{louis:lens:2013}.
On large scales ($\ell < 500$), this tells the same story as the maps did:
The larger the scale, the more slowly it converges, with the constant correlation
preconditioner being about 3 times faster than the binned one. Somewhat surprisingly,
a similar phenomenon occurs at the small scales. For $\ell > 2000$, higher
$\ell$ results in slower convergence, and this is especially prominent
for the EE and BB power spectra. On all scales, however, the constant correlation
preconditioner appears to converge several times faster than the binned one.

In order to quantify the convergence more precisely, we consider the
time at which the absolute error in a given multipole-bin reaches
0.1 times cosmic variance in that bin. While somewhat arbitrary, this
choice ensures that CG errors are guaranteed to be sub-dominant
in the power spectrum, regardless of the noise properties of the
actual experiment.
Figure~\ref{fig:ps_conv} shows the convergence of TT, EE and BB for both
preconditioners for a typical multipole-bin. The overall trend for each component
is an initial rapid fall followed by a slower decay, with both being
significantly faster for the new preconditioner, particularly for the
polarization spectra.

\begin{figure}[htb]
	\centering
	\includegraphics[width=\columnwidth]{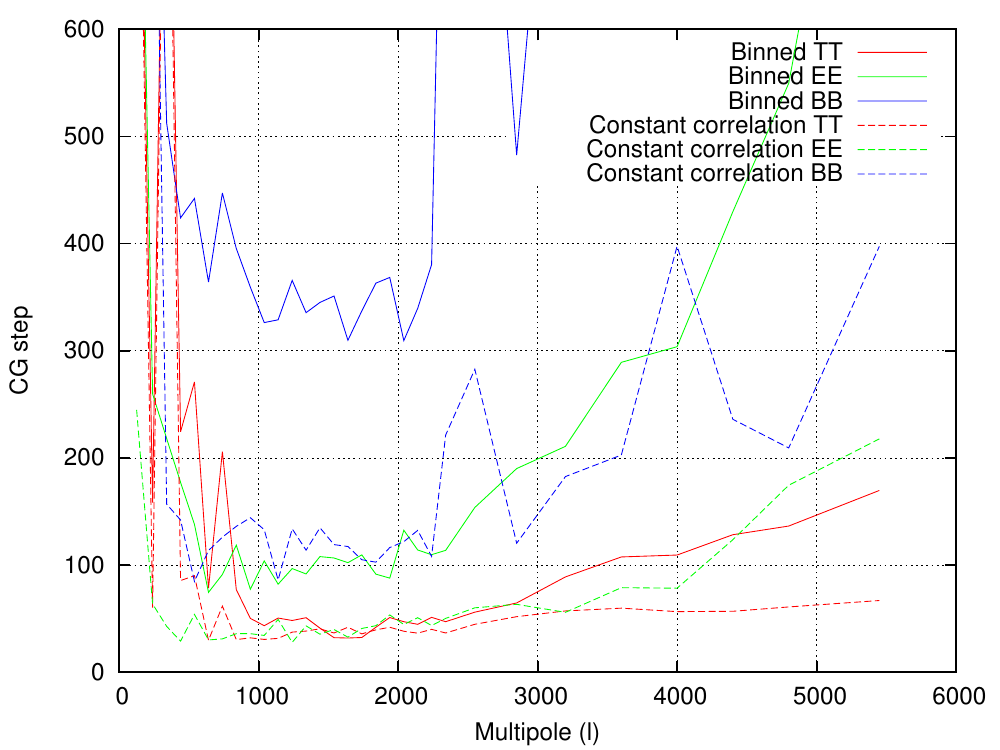}
	\caption{The number of CG steps needed for each multipole-bin
		to converge to 0.1 times cosmic variance, for each of TT (red), EE (green) and BB (blue)
		for the binned (solid) and constant correlation (dashed)
		preconditioners.}
	\label{fig:all_conv}
\end{figure}

We found that binning the errors in bins of $\sim50$ conjugate gradients steps
and using linear interpolation between these bins resulted in a robust estimate
of when each spectrum bin reaches the convergence criterion. The resulting
convergence times can be seen in figure~\ref{fig:all_conv},
and confirm our earlier finding that the largest and smallest scales
converge more slowly. The figure also highlights how much trouble the binned
preconditioner has with the EE and especially BB spectra, where it
performs much more poorly relative the constant correlation preconditioner
than we see for the TT spectrum. We speculate that this is due to
the X-shaped correlation structure introduced
by our scanning pattern. In binned maps, which ignore the correlations,
this introduces spurious X-shaped patterns in both Q and U,
corresponding to spurious signal in both E and B.
With
the constant correlation preconditioner, these are partially corrected
because some of the correlation structure is taken into account.

\begin{figure}[htb]
	\centering
	\includegraphics[width=\columnwidth]{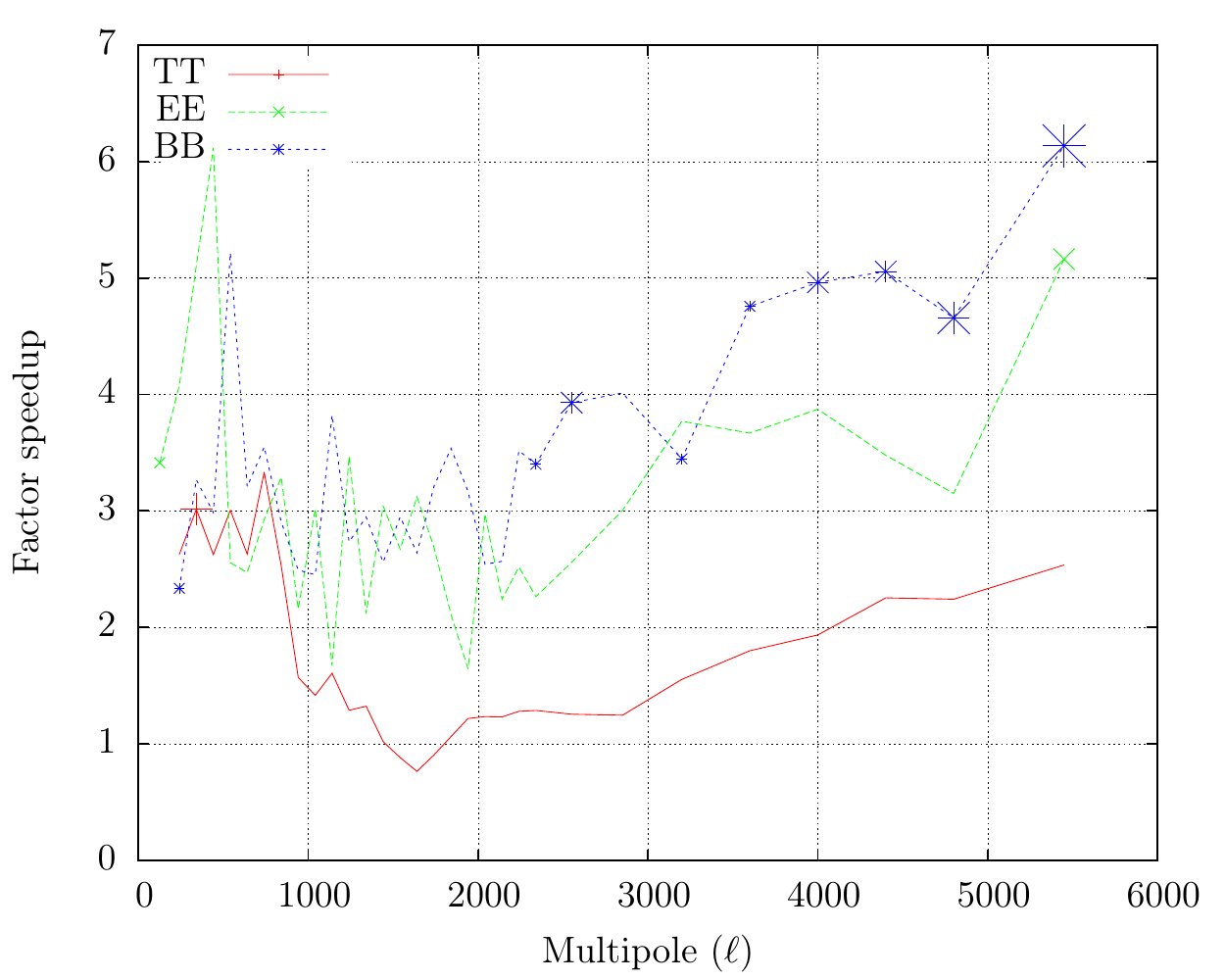}
	\caption{The ratio of the convergence times to 0.1 times cosmic
		variance for the binned and
		constant correlation preconditioners, per multipole bin for
		each of TT, EE and BB. For TT, the speedup is typically between 20\%
		and 200\%, while for BB the speedup ranges from 200\% at large scales
		to about 400\% at small scales, with EE being intermediate. Large
		points here indicate cases where the binned preconditioner did not
		converge to 0.1 times cosmic variance in time. For these points, the comparison
		was performed at the lowest threshold where both converged.}
	\label{fig:speedup}
\end{figure}

We summarize the performance characteristics of the new
preconditioner in figure~\ref{fig:speedup}, which shows
its relative speed gain vs. the baseline binned preconditioner.
Depending on $\ell$, we have a speedup
ranging from 20\% to 200\% for TT, from 150\% to 300\% for EE,
and from 200\% to 400\% for BB, with the greatest relative
improvement happening at the scales that converge most slowly.

\section{Summary}
The structure of a CMB map's pixel covariance matrix is
determined by the noise properties of the time-ordered data
and the scanning pattern of the telescope.
For constant elevation drift scans like
those employed by ACT, SPT and POLARBEAR, this results in an approximately circulant
covariance. We have developed a new preconditioner for conjugate
gradient solutions of the map-making equation which exploit this
property by deconvolving the correlations in harmonic space,
an operation which is very cheap due to the Fourier representation
of a circulant matrices being diagonal.

For a realistic scanning pattern and noise model
the preconditioner results a speedup of 20\% to 200\% for
temperature and 150\% to 400\% for polarization compared to
a binned preconditioner.

Convergence speed might potentially be further improved by
allowing the correlation pattern to change slowly across the map,
for example by decomposing the map into overlapping tiles, and
applying the constant correlation preconditioner separately to
each tile, followed by a merging operation. Our preliminary
attempts at such a tiled preconditioner have however not
been able to beat the simple constant correlation approximation
presented here.

\begin{acknowledgments}
The authors would like to thank Jo Dunkley and Johannes Noller for useful
discussion and suggestions, and Jon Sievers for testing the preconditioner in another
map-maker. We also thank the ACT collaboration for access to internal
ACT data used in the simulations.
Computations were performed on the gpc supercomputer at
the SciNet HPC Consortium. SciNet is funded by: the Canada Foundation for
Innovation under the auspices of Compute Canada; the Government of Ontario;
Ontario Research Fund - Research Excellence; and the University of Toronto.
SN and TL are supported by ERC grant 259505.
\end{acknowledgments}

\bibliography{refs}
\bibliographystyle{apj}

\end{document}